\documentclass[12pt,letterpaper,english,aps,superscriptaddress,groupedaddress,showpacs,showkeys]{revtex4}
\usepackage[T1]{fontenc}
\usepackage[latin9]{inputenc}
\usepackage{babel}
\usepackage{amsmath}
\usepackage{amssymb}
\usepackage{esint}
\usepackage[unicode=true]
 {hyperref}
\usepackage{breakurl}

\makeatletter

\@ifundefined{textcolor}{}
{%
 \definecolor{BLACK}{gray}{0}
 \definecolor{WHITE}{gray}{1}
 \definecolor{RED}{rgb}{1,0,0}
 \definecolor{GREEN}{rgb}{0,1,0}
 \definecolor{BLUE}{rgb}{0,0,1}
 \definecolor{CYAN}{cmyk}{1,0,0,0}
 \definecolor{MAGENTA}{cmyk}{0,1,0,0}
 \definecolor{YELLOW}{cmyk}{0,0,1,0}
}

\usepackage{dcolumn}
\usepackage{bm}
\usepackage{amsfonts}

\makeatother

\begin{document}

\title{On real solutions of the Dirac equation for a one-dimensional Majorana
particle}

\author{Salvatore De Vincenzo}

\email{[salvatore.devincenzo@ucv.ve]}

\selectlanguage{english}%

\thanks{I would like to dedicate this paper to the memory of my beloved father
Carmine De Vincenzo Di Fresca, who passed away unexpectedly on March
16, 2018. That day something of me also died.}

\affiliation{Escuela de F\'{\i}sica, Facultad de Ciencias, Universidad Central de Venezuela,
A.P. 47145, Caracas 1041-A, Venezuela.}

\date{July 23, 2019}
\begin{abstract}
We construct general solutions of the time-dependent Dirac equation
in (1+1) dimensions with a Lorentz scalar potential, subject to the
so-called Majorana condition, in the Majorana representation. In this
situation, these solutions are real-valued and describe a one-dimensional
Majorana single particle. We specifically obtain solutions for the
following cases: a Majorana particle at rest inside a box, a free
(i.e., in a penetrable box with the periodic boundary condition),
in an impenetrable box with no potential (here we only have four boundary
conditions), and in a linear potential. All these problems are treated
in a very detailed and systematic way. In addition, we obtain and
discuss various results related to real wave functions. Finally, we
also wish to point out that, in choosing the Majorana representation,
the solutions of the Dirac equation with a Lorentz scalar potential
can be chosen to be real but do not need to be real. In fact, complex
solutions for this equation can also be obtained. Thus, a Majorana
particle cannot be described only with the Dirac equation in the Majorana
representation without explicitly imposing the Majorana condition.
\end{abstract}



\maketitle
\section{Introduction}

In 1937, Majorana posed the question of whether a particle
with a spin of 1/2 could be its own antiparticle \cite{RefA}. Essentially,
Majorana noted that if the gamma matrices present in the free Dirac
equation 
\begin{equation}
\left(\mathrm{i}\hat{\gamma}^{\mu}\partial_{\mu}-\frac{\mathrm{m}c}{\hbar}\hat{1}_{4}\right)\Psi=0
\end{equation}
were forced to satisfy the condition $(\mathrm{i}\hat{\gamma}^{\mu})^{*}=\mathrm{i}\hat{\gamma}^{\mu}$
($\hat{1}_{4}$ is the $4\times4$ identity matrix), then Eq. (1)
could have real-valued solutions. Then, Majorana found for the first
time a set of matrices that satisfy this condition (this is the so-called
Majorana representation). In other words, in the Majorana representation,
the free complex Dirac operator (i.e., the operator acting on $\Psi$
in Eq. (1)) becomes a real Dirac operator; therefore, one can obtain
real-valued solutions for the Dirac equation. Actually, in choosing
the matrices in the Majorana representation, the solutions of the
free Dirac equation could be chosen to be real but do not need to
be real. It is found that these results are still valid when a (real)
Lorentz scalar potential is included into the free Dirac equation.
In fact, no other Lorentz potential can be added; for example, in
(1+1) dimensions, neither a Lorentz two-vector nor a Lorentz pseudoscalar
potential can be added \cite{RefB}. On a side note, the Klein-Fock-Gordon
and Maxwell equations are also real equations; therefore, they can
accommodate real solutions. These equations, however, can also have
complex solutions.

The particle described by the real-valued solution of the Dirac equation
is the Majorana particle, which would be, in fact, a massive fermion
that is its own antiparticle (at least in (3+1) dimensions); hence,
the particle must be electrically neutral \cite{RefA,RefC}. In (1+1)
dimensions, it is understood that this particle is the one-dimensional
Majorana particle. To date, no elementary particle has been identified
as a Majorana particle. Among the known spin-1/2 particles, only neutrinos
could be of a Majorana nature \cite{RefD}. However, a different type
of Majorana fermion have recently emerged within condensed matter
systems as exotic quasiparticle excitations that are their own antiparticles;
for instance, see \cite{RefD,RefE,RefF,RefG} and the references within. 

Thus, in the Majorana representation, the equation that describes
a Majorana particle is the Dirac equation together with the reality
condition of the wave function
\begin{equation}
\Psi=\Psi^{*}.
\end{equation}
Precisely, in the Majorana representation, $\Psi^{*}$ is the charge-conjugate
wave function of $\Psi$, $\Psi_{C}$ \cite{RefB,RefG}; therefore,
the reality condition given by Eq. (2) expresses the invariance of
$\Psi$ under the charge-conjugation operation, i.e., 
\begin{equation}
\Psi=\Psi_{C}.
\end{equation}
The latter relation is what defines a Majorana particle in any representation
\cite{RefH} and is called the Majorana condition. Certainly, depending
on the representation, the equation for the Majorana particle is a
real or a complex system of coupled equations or a complex single
equation for a single component of the Dirac wave function \cite{RefG,RefB}.
Moreover, it is important to remember that $\Psi_{C}$ has the same
transformation properties as $\Psi$ under proper Lorentz transformations
\cite{RefH}; hence, the Majorana condition is Lorentz covariant. 

Traditionally, the Dirac wave function $\Psi$ has been physically
associated with the Dirac particle, and its charge-conjugate wave
function $\Psi_{C}$ has been associated with its antiparticle. This
is because charge conjugation is essentially an operation that changes
a particle into its antiparticle (and vice versa). However, in a single-particle
theory, if $\Psi$ describes the particle's state with positive energy,
for example, in a given Lorentz scalar potential, then $\Psi_{C}$
describes the particle's state (not its antiparticle's state) with
negative energy, in the same potential (and whose absence appears
as the presence of a physical antiparticle with positive energy, according
to the so-called Dirac hole theory). Apropos of this, in the Majorana
representation, $\Psi^{*}$ can describe that particle's state with
negative energy, but the Majorana condition continues to give us a
real solution for the time-dependent Dirac equation. In any case,
one can always invoke the hole theory to obtain a physical picture
of the negative energy states (see, for example, pp. 131-134 in Ref.
\cite{RefI}). However, as is well known, the latter takes us out
of the Dirac single-particle theory itself (recall the unobservable
infinite Dirac sea of negative energy particles, the so-called ``vacuum''
state). 

To recap, a Majorana particle cannot be described only with the Dirac
equation in the Majorana representation without imposing the Majorana
condition. In the present paper, we want to emphasize this point and
highlight the importance of the Majorana condition as a necessary
condition to characterize a Majorana particle. We also consider that
this task is absolutely pertinent; in fact, we have seen a recently
published article that incorrectly assumes that a Majorana particle
(in this case, subject to the linear Lorentz scalar potential) can
be described just with the Dirac equation in the Majorana representation,
without imposing the Majorana condition \cite{RefJ}. 

The paper is organized as follows. The results are given in section
II. First, we present the wave equation that describes a (one-dimensional)
Majorana particle, namely, the time-dependent Dirac equation (in (1+1)
dimensions) with a Lorentz scalar potential with its solutions restricted
by the Majorana condition. We use the Majorana representation; thus,
the Majorana condition will ensure that the Dirac wave function is
real. Then, we construct, step-by-step, the real general solutions
of this equation in the following rather simple cases: a Majorana
particle at rest inside a box (with no potential), a free (i.e., in
a penetrable box with a periodic boundary condition), in an impenetrable
box with no potential (here we only have four possible boundary conditions),
and in a linear potential. All these problems are treated in a very
detailed and systematic way. In addition, throughout this section
we obtain and discuss various results, some of which are related to
real wave functions. For example, we note that a typical wave equation
of first order in the time derivative can have real solutions, but
then the corresponding Hamiltonian must be purely imaginary. If the
latter condition is fulfilled, and additionally if this operator is
formally self-adjoint, then its mean value in a real-valued state
always vanishes. The latter result can also be extended to the momentum
operator. Finally, we wish to note that in this article we are particularly
careful when dealing with operators. For example, we respect the essential
difference that exists between a Hermitian operator (or a formally
self-adjoint operator) and a self-adjoint operator. Because in quantum
mechanics observables are always represented by self-adjoint operators,
we take into account only the boundary conditions that are included
in the domains of (unbounded) self-adjoint operators. Lastly, the
conclusions are presented in section III. 

\section{The Majorana particle}

The Dirac equation describing first quantized Dirac particles
(in (1+1) dimensions) in a real-valued Lorentz scalar potential $\mathcal{S}=\mathcal{S}(x)$
has the form \cite{RefB}
\begin{equation}
\left[\,\mathrm{i}\hat{\gamma}^{\mu}\partial_{\mu}-\frac{1}{\hbar c}(\mathcal{S}+\mathrm{m}c^{2})\hat{1}_{2}\,\right]\Psi=0.
\end{equation}
The Dirac single-particle two-component wave function $\Psi=\left[\,\phi_{1}\;\phi_{2}\,\right]^{\mathrm{T}}$
is generally complex-valued (the symbol \rm{T} represents the transpose
of a matrix). The Dirac gamma matrices $\hat{\gamma}^{\mu}$, with
$\mu=0,1$, satisfy $\hat{\gamma}^{\mu}\hat{\gamma}^{\nu}+\hat{\gamma}^{\nu}\hat{\gamma}^{\mu}=2g^{\mu\nu}\hat{1}_{2}$,
where $g^{\mu\nu}=\mathrm{diag}(1,-1)$ ($\hat{1}_{2}$ is the $2\times2$
identity matrix). In the Majorana representation, the Dirac gamma
matrices satisfy $(\mathrm{i}\hat{\gamma}^{\mu})^{*}=\mathrm{i}\hat{\gamma}^{\mu}$,
and the operator acting on $\Psi$ in Eq. (4) is real (the raised
asterisk $^{*}$ denotes a complex conjugate). In the latter case,
the solutions of the Dirac equation (4) can be chosen to be real,
but clearly they do not need to be real. For the Majorana representation,
we choose the following matrices:
\begin{equation}
\hat{\gamma}^{0}=\hat{\sigma}_{y}\,,\quad\hat{\gamma}^{1}=-\mathrm{i}\hat{\sigma}_{z},
\end{equation}
where $\hat{\sigma}_{y}$ and $\hat{\sigma}_{z}$ are the Pauli matrices.
Obviously, the matrices $\hat{\gamma}^{0}\,'=\hat{\sigma}_{y}$ and
$\hat{\gamma}^{1}\,'=+\mathrm{i}\hat{\sigma}_{z}$, for example, can
also be considered to be a Majorana representation (incidentally,
the latter choice was used in Refs. \cite{RefJ,RefK}). These two
Majorana representations are related by $\hat{\gamma}^{\mu}\,'=\hat{\sigma}_{y}\,\hat{\gamma}^{\mu}\,\hat{\sigma}_{y}$
and $\left[\,\phi_{1}'\;\phi_{2}'\,\right]^{\mathrm{T}}=\hat{\sigma}_{y}\left[\,\phi_{1}\;\phi_{2}\,\right]^{\mathrm{T}}$. 

The Dirac equation (4) also describes Majorana particles if, in addition,
$\Psi$ complies, in any representation, with so-called Majorana condition
\begin{equation}
\Psi=\Psi_{C},
\end{equation}
where $\Psi_{C}$, the charge-conjugate wave function, also satisfies
Eq. (4) (remember that the scalar potential would be independent of
the charge of the particle \cite{RefB}), but this implies that
\begin{equation}
\hat{S}_{C}\,(\mathrm{i}\hat{\gamma}^{\mu})^{*}(\hat{S}_{C})^{-1}=\mathrm{i}\hat{\gamma}^{\mu}\,,\quad\mathrm{where}\quad\Psi_{C}\equiv\hat{S}_{C}\,\Psi^{*},
\end{equation}
and the matrix $\hat{S}_{C}$ can be chosen to be unitary (up to a
phase factor). In the Majorana representation, we have $\hat{S}_{C}=\hat{1}_{2}$;
therefore, $(\mathrm{i}\hat{\gamma}^{\mu})^{*}=\mathrm{i}\hat{\gamma}^{\mu}$
(as expected), and also $\Psi_{C}=\Psi^{*}$ (by virtue of Eq. (7)).
Consequently, the Majorana condition takes the form
\begin{equation}
\Psi=\Psi^{*}\,(\,\Leftrightarrow\,\phi_{1}=\phi_{1}^{*}\:\,\mathrm{and}\:\,\phi_{2}=\phi_{2}^{*}\,),
\end{equation}
i.e., the entire Dirac wave function must be real, and it becomes
the Majorana wave function. 

It is worth mentioning that the Majorana condition can also be written
as $\Psi=\exp(\mathrm{i}\theta)\Psi_{C}$, where $\theta$ is an arbitrary
phase. We could call this condition the generalized Majorana condition,
and it remains covariant under a Lorentz transformation, as it must
\cite{RefH}. Certainly, if this version of the Majorana condition
is explicitly used, every solution $\Psi$ that describes a Majorana
particle will depend on the phase $\theta$. We can always choose
$\theta=0$, but the freedom to choose a different value for the angle
$\theta$ could be convenient. Lastly, it is important to note that
this generalized Majorana condition also leads to real solutions in
the Majorana representation. Moreover, this condition leads to the
Majorana condition in the form given in Eq. (6) for a wave function
$\tilde{\Psi}$ that is physically indistinguishable from $\Psi$.
In effect, if the generalized Majorana condition is verified, then
the following relation is verified: $\tilde{\Psi}=\hat{S}_{C}\,\tilde{\Psi}^{*}\,(\equiv\tilde{\Psi}_{C})$,
where $\tilde{\Psi}\equiv\exp(-\mathrm{i}\theta/2)\Psi$. 

The Dirac equation (4) in its Hamiltonian form is 
\begin{equation}
\left(\mathrm{i}\hbar\hat{1}_{2}\,\frac{\partial}{\partial t}-\hat{\mathrm{h}}\right)\Psi=0,
\end{equation}
where 
\begin{equation}
\hat{\mathrm{h}}=c\,\hat{\alpha}\hat{\mathrm{p}}+(\mathcal{S}+\mathrm{m}c^{2})\hat{\beta}
\end{equation}
is the formally self-adjoint, or Hermitian, Hamiltonian operator,
i.e., $\hat{\mathrm{h}}=\hat{\mathrm{h}}^{\dagger}$ (in the present
paper, the symbol $\dagger$ denotes the Hermitian conjugate, or the
adjoint, of a matrix and an operator). The Dirac matrices are $\hat{\alpha}=\hat{\gamma}^{0}\hat{\gamma}^{1}$
and $\hat{\beta}=\hat{\gamma}^{0}$, and $\hat{\mathrm{p}}=-\mathrm{i}\hbar\hat{1}_{2}\,\partial/\partial x=\hat{\mathrm{p}}^{\dagger}$
is the formally self-adjoint, or Hermitian, Dirac momentum operator
(which is, in the end, a $2\times2$ matrix). In any Majorana representation,
we have that $(\mathrm{i}\hat{\gamma}^{\mu})^{*}=\mathrm{i}\hat{\gamma}^{\mu}$;
therefore, $\hat{\alpha}=\hat{\alpha}^{*}$ and $\hat{\beta}=-\hat{\beta}^{*}$.
The latter implies that
\begin{equation}
\hat{\mathrm{h}}=-\hat{\mathrm{h}}^{*},
\end{equation}
and as expected, the operator acting on $\Psi$ in Eq. (9) is real.
Thus, Eq. (11) defines a Dirac Hamiltonian in the Majorana representation.
In general, if one has a dynamic equation of the form $(\mathrm{i}\partial_{t}-\hat{\mathrm{H}})\Phi=0$
($\partial_{t}\equiv\partial/\partial t$), the operator acting on
$\Phi$ is real if $\hat{\mathrm{H}}=-\hat{\mathrm{H}}^{*}$, i.e.,
if $\hat{\mathrm{H}}$ is purely imaginary. For instance, this result
tell us that the time-dependent Schr\"odinger equation cannot have real
solutions, i.e., it cannot have solutions \textit{\`a la} Majorana.

In the Majorana representation that we have chosen in the present
article, we have that $\hat{\alpha}=\hat{\sigma}_{x}$ and $\hat{\beta}=\hat{\sigma}_{y}$.
By substituting these matrices into Eq. (10) and substituting the
result obtained into Eq. (9), we obtain the following real system
of two coupled equations for the functions $\phi_{1}$ and $\phi_{2}$:
\begin{equation}
\left[\begin{array}{cc}
\mathrm{i}\hbar\,\frac{\partial}{\partial t} & \mathrm{i}\hbar c\,\frac{\partial}{\partial x}+\mathrm{i}(\mathcal{S}+\mathrm{m}c^{2})\\
\mathrm{i}\hbar c\,\frac{\partial}{\partial x}-\mathrm{i}(\mathcal{S}+\mathrm{m}c^{2}) & \mathrm{i}\hbar\,\frac{\partial}{\partial t}
\end{array}\right]\left[\begin{array}{c}
\phi_{1}\\
\phi_{2}
\end{array}\right]=\left[\begin{array}{c}
0\\
0
\end{array}\right]
\end{equation}
(of course, you can cancel the imaginary number!). We can also prove
that each component of $\Psi$ satisfies an equation of the Klein-Fock-Gordon
type, namely,
\begin{equation}
\left[\,\frac{1}{c^{2}}\frac{\partial^{2}}{\partial t^{2}}-\frac{\partial^{2}}{\partial x^{2}}+\frac{(-1)^{j-1}}{\hbar c}\frac{\mathrm{d}\mathcal{S}}{\mathrm{d}x}+\frac{(\mathcal{S}+\mathrm{m}c^{2})^{2}}{\hbar^{2}c^{2}}\,\right]\phi_{j}=0\,,\quad j=1,2.
\end{equation}
The latter two equations are only slightly different, but in the free
case ($\mathcal{S}=\mathrm{const}$), the equations are the same.
In the latter situation, Eq. (13) is precisely the free Klein-Fock-Gordon
equation with mass $\mathrm{m}c^{2}+\mathrm{const}$. 

Obviously, the system of equations (12) is real because it is precisely
the Dirac equation in the Majorana representation. If we explicitly
demand that the solutions $\phi_{1}$ and $\phi_{2}$ are real-valued,
then $\Psi=\left[\,\phi_{1}\;\phi_{2}\,\right]^{\mathrm{T}}$ describes
a Majorana particle, and the Majorana condition $\Psi=\Psi_{C}\,(=\Psi^{*})$
is satisfied. Complex solutions for Eq. (12) can also be obtained
\cite{RefJ}, but these do not describe a Majorana particle. 

\subsection{Majorana particle at rest inside a box}

This case will allow us to present the essential ideas of
the subject in an immediate and simple way. To describe a (one-dimensional)
Majorana particle at rest inside a box with no potential ($\mathrm{p}=\mathcal{S}=0$),
the first thing we must do is solve the following Dirac equation:
\begin{equation}
\mathrm{i}\hbar\hat{1}_{2}\,\frac{\mathrm{d}}{\mathrm{d}t}\Psi=\hat{\mathrm{h}}\Psi=\mathrm{m}c^{2}\hat{\sigma}_{y}\Psi
\end{equation}
(see Eqs. (9) and (10)). In this case, $\hat{\mathrm{h}}$ is simply
a $2\times2$ Hermitian constant matrix, i.e., $\hat{\mathrm{h}}=\hat{\mathrm{h}}^{\dagger}$;
in fact, $\hat{\mathrm{h}}$ is also self-adjoint, because it is a
bounded operator. It could be considered that this Dirac Hamiltonian
operator acts on two-component column vectors $\Psi$ that belong
to the Hilbert space of the square integrable functions, $\mathcal{H}=\mathcal{L}^{2}(\Omega)\oplus\mathcal{L}^{2}(\Omega)$
(a direct sum), where $\Omega$ is a region in $\mathbb{R}$ of width
$L$ (a box) that can be arbitrarily large. In general, $\Psi$ is
a function of $x$ with values in $\mathbb{C}^{2}$ (the two-dimensional
complex linear space); hence, the Hilbert space could also be written
as $\mathcal{H}=\mathcal{L}^{2}(\Omega)\otimes\mathbb{C}^{2}$ (a
tensor product). The scalar product in $\mathcal{H}$ is denoted by
$\langle\Phi,\chi\rangle\equiv\int_{\Omega}\,\mathrm{d}x\,\Phi^{\dagger}\chi$
(the integrand is the $\mathbb{C}^{2}$-scalar product of the $\mathbb{C}^{2}$-vectors
$\Phi$ and $\chi$), and the norm is $\left\Vert \,\Phi\,\right\Vert \equiv\sqrt{\langle\Phi,\Phi\rangle}$
\cite{RefL}. The eigenfunctions of $\hat{\mathrm{h}}$ are denoted
by $\psi_{\mathrm{p}=0}^{(+)}\equiv\psi_{0}^{(+)}$ and $\psi_{\mathrm{p}=0}^{(-)}\equiv\psi_{0}^{(-)}$
for the eigenvalues $E=+E_{\mathrm{p}=0}\equiv+E_{0}=+\mathrm{m}c^{2}$
(positive-energy) and $E=-E_{\mathrm{p}=0}\equiv-E_{0}$ (negative-energy),
respectively:
\begin{equation}
\hat{\mathrm{h}}\,\psi_{0}^{(+)}=+\mathrm{m}c^{2}\psi_{0}^{(+)}\,,\quad\hat{\mathrm{h}}\,\psi_{0}^{(-)}=-\mathrm{m}c^{2}\psi_{0}^{(-)}.
\end{equation}
Explicitly, these eigenfunctions can be written as follows
\begin{equation}
\psi_{0}^{(+)}\equiv\sqrt{\frac{1}{L}}\, u(0)=\sqrt{\frac{1}{2L}}\left[\begin{array}{c}
1\\
\mathrm{i}
\end{array}\right]
\end{equation}
and 
\begin{equation}
\psi_{0}^{(-)}=(\psi_{0}^{(+)})_{C}=(\psi_{0}^{(+)})^{*}=\sqrt{\frac{1}{L}}\,(u(0))^{*}\equiv\sqrt{\frac{1}{L}}\, v(0)=\sqrt{\frac{1}{2L}}\left[\begin{array}{c}
1\\
-\mathrm{i}
\end{array}\right].
\end{equation}
Moreover, they satisfy the following relations of normalization and
orthogonality, as expected:
\begin{equation}
\langle\psi_{0}^{(+)},\psi_{0}^{(+)}\rangle=\langle\psi_{0}^{(-)},\psi_{0}^{(-)}\rangle=\,1\,,\quad\langle\psi_{0}^{(+)},\psi_{0}^{(-)}\rangle=0,
\end{equation}
but the following $\mathbb{C}^{2}$-orthogonality and normalization
conditions for the two-component column vectors $u(0)$ and $v(0)$
(that are charge-conjugates of each other) must also be satisfied:
\begin{equation}
(u(0))^{\dagger}u(0)=(v(0))^{\dagger}v(0)\equiv\frac{E_{0}}{\mathrm{m}c^{2}}=1\,,\quad(u(0))^{\dagger}v(0)=0.
\end{equation}

The most general solution of Eq. (14) at $t=0$, i.e., the initial
state, can be written as
\begin{equation}
\Psi(t=0)\equiv\Psi(0)=c_{0}^{(+)}(0)\,\psi_{0}^{(+)}+c_{0}^{(-)}(0)\,\psi_{0}^{(-)},
\end{equation}
where $c_{0}^{(\pm)}(0)\equiv c_{\mathrm{p}=0}^{(\pm)}(t=0)$ are
two arbitrary complex constants. But if $\Psi(0)$ is required to
be normalized, then the following constraint is verified:
\begin{equation}
\left|\, c_{0}^{(+)}(0)\,\right|^{2}+\left|\, c_{0}^{(-)}(0)\,\right|^{2}=1.
\end{equation}
Because, in the end, we want to describe a Majorana particle at rest,
we impose upon the wave function $\Psi(0)$ the Majorana condition,
i.e., $\Psi(0)=\Psi_{C}(0)=(\Psi(0))^{*}$. The latter implies that
$c_{0}^{(-)}(0)=(c_{0}^{(+)}(0))^{*}$; therefore, the most general
initial state for the Majorana particle at rest can be written as
follows
\begin{equation}
\Psi(0)=c_{0}^{(+)}(0)\,\psi_{0}^{(+)}+\mathrm{c.c.}\,,\quad\left|\, c_{0}^{(+)}(0)\,\right|=\frac{1}{\sqrt{2}},
\end{equation}
where, henceforth, $\mathrm{c.c.}$ means a complex conjugate, i.e.,
the complex conjugate of the expression to the left of the symbol
$\mathrm{c.c.}$.

The state of the system at any subsequent time $t$ is simply
\begin{equation}
\Psi(t)=\hat{U}(t)\Psi(0),
\end{equation}
where 
\begin{equation}
\hat{U}(t)=\exp\left(-\frac{\mathrm{i}}{\hbar}\, t\,\hat{\mathrm{h}}\right)=\sum_{k=0}^{\infty}\frac{1}{k!}\left(-\frac{\mathrm{i}}{\hbar}\, t\,\hat{\mathrm{h}}\right)^{k}=\exp\left(-\mathrm{i}\,\frac{\mathrm{m}c^{2}}{\hbar}\, t\,\hat{\sigma}_{y}\right)
\end{equation}
is the evolution operator. Note that, in addition to being unitary,
this operator is real in this case. Thus, because $\Psi(0)$ is real,
then $\Psi(t)$ will also be real (in the same way, if $\Psi(0)$
would have been complex, then $\Psi(t)$ would also be complex). It
is worthwhile to mention that, because $\hat{\mathrm{h}}$ is a matrix,
$\hat{U}(t)$ admits correctly the power series expansion in Eq. (24).
Precisely, by using the latter expansion when applying $\hat{U}(t)$
to $\Psi(0)$, we finally obtain from Eq. (23) the Majorana wave function
at the instant $t$ 
\begin{equation}
\Psi(t)=c_{0}^{(+)}(0)\,\psi_{0}^{(+)}\exp\left(-\mathrm{i}\,\frac{\mathrm{m}c^{2}}{\hbar}\, t\right)+\mathrm{c.c.},
\end{equation}
with $\left|\, c_{0}^{(+)}(0)\,\right|=\frac{1}{\sqrt{2}}$. Clearly,
the real general state in Eq. (25) is an equal-weight superposition
of the positive and negative energy complex eigenstates, i.e., of
two complex stationary states, for all $t$. Incidentally, if we had
used the Majorana condition in its generalized form, namely, $\Psi=\exp(\mathrm{i}\theta)\Psi_{C}$,
then $\Psi(0)$ in Eq. (22) and $\Psi(t)$ in Eq. (25) would have
depended of $\theta$. Specifically, we would have to multiply the
c.c. term in those expressions by the phase factor $\exp(\mathrm{i}\theta)$. 

It can be demonstrated that the mean value of $\hat{\mathrm{h}}$
in the real normalized state $\Psi(t)$ vanishes. In effect, first
of all, we have that
\begin{equation}
\langle\hat{\mathrm{h}}\rangle_{\Psi}\equiv\langle\Psi,\hat{\mathrm{h}}\Psi\rangle=-\langle\Psi,\hat{\mathrm{h}}^{*}\Psi\rangle=-\langle\Psi^{*},\hat{\mathrm{h}}^{*}\Psi^{*}\rangle=-\langle\Psi,\hat{\mathrm{h}}\Psi\rangle^{*}=-\langle\hat{\mathrm{h}}\rangle_{\Psi}^{*},
\end{equation}
where we have made use of Eq. (11). Second, just because $\hat{\mathrm{h}}$
is a Hermitian matrix, the following expected result is verified:
\begin{equation}
\langle\hat{\mathrm{h}}\rangle_{\Psi}\equiv\langle\Psi,\hat{\mathrm{h}}\Psi\rangle=\int_{\Omega}\,\mathrm{d}x\,\Psi^{\dagger}\,\hat{\mathrm{h}}\,\Psi=\int_{\Omega}\,\mathrm{d}x\,(\hat{\mathrm{h}}\Psi)^{\dagger}\,\Psi=\langle\hat{\mathrm{h}}\Psi,\Psi\rangle=\langle\Psi,\hat{\mathrm{h}}\Psi\rangle^{*}=\langle\hat{\mathrm{h}}\rangle_{\Psi}^{*},
\end{equation}
where in the penultimate step, we made use of the scalar product property
$\langle\chi,\Phi\rangle=\langle\Phi,\chi\rangle^{*}$. Finally, from
the results given in Eqs. (26) and (27), we obtain 
\begin{equation}
\langle\hat{\mathrm{h}}\rangle_{\Psi}=0.
\end{equation}

Clearly, the latter is a general result; in fact, the mean value of
a formally self-adjoint operator $\hat{\mathrm{H}}=\hat{\mathrm{H}}^{\dagger}$
(not just a Hermitian matrix), which also satisfies the equality $\hat{\mathrm{H}}=-\hat{\mathrm{H}}^{*}$,
in a real-valued state, always vanishes. This is simply because the
former condition implies that $\langle\hat{\mathrm{H}}\rangle_{\Psi}=\langle\hat{\mathrm{H}}\rangle_{\Psi}^{*}$,
and the latter condition implies $\langle\hat{\mathrm{H}}\rangle_{\Psi\in\mathbb{R}}=-\langle\hat{\mathrm{H}}\rangle_{\Psi\in\mathbb{R}}^{*}$;
thus, $\langle\hat{\mathrm{H}}\rangle_{\Psi\in\mathbb{R}}=0$. For
instance, the Dirac momentum operator $\hat{\mathrm{p}}=-\mathrm{i}\hbar\hat{1}_{2}\,\partial/\partial x$
satisfies the relation $\hat{\mathrm{p}}=-\hat{\mathrm{p}}^{*}$,
and because it is also a formally self-adjoint operator, we can obtain
the result $\langle\hat{\mathrm{p}}\rangle_{\Psi\in\mathbb{R}}=0$.
On the other hand, the Dirac or standard velocity operator, $\hat{\mathrm{v}}\equiv c\hat{\alpha}$,
is a Hermitian matrix but satisfies the relation $\hat{\mathrm{v}}=+\hat{\mathrm{v}}^{*}$
(this is always so in the Majorana representation); therefore, the
mean value $\langle\hat{\mathrm{v}}\rangle_{\Psi\in\mathbb{R}}$ does
not have to be zero, although it is real-valued \cite{RefM}. Likewise,
the so-called classical velocity operator, $\hat{\mathrm{v}}_{\mathrm{cl}}\equiv c^{2}\,\hat{\mathrm{p}}\,\hat{\mathrm{h}}^{-1}$
(that corresponds to the formula of classical relativistic mechanics
that gives the velocity as a function of momentum and energy) \cite{RefL}
is a real and formally self-adjoint operator, i.e., $\hat{\mathrm{v}}_{\mathrm{cl}}=\hat{\mathrm{v}}_{\mathrm{cl}}^{*}$
and $\hat{\mathrm{v}}_{\mathrm{cl}}=\hat{\mathrm{v}}_{\mathrm{cl}}^{\dagger}$;
therefore, its (real) expectation value in a real state, $\langle\hat{\mathrm{v}}_{\mathrm{cl}}\rangle_{\Psi\in\mathbb{R}}$,
does not have to be zero. 

It is worth mentioning that in the present case (and approximately
even when $\mathrm{p}\approx0$, i.e., $\mathrm{p}\ll\mathrm{m}c$),
each component of $\Psi(t)$ satisfies the equation of a harmonic
oscillator with frequency $\omega\equiv\mathrm{m}c^{2}/\hbar$, namely
(see Eq. (13)),
\begin{equation}
\left(\,\frac{\mathrm{d}^{2}}{\mathrm{d}t^{2}}+\omega^{2}\,\right)\phi_{j}=0\,,\quad j=1,2.
\end{equation}
Solving these equations and obviously respecting the connections between
$\phi_{1}(t)$ and $\phi_{2}(t)$ from the Dirac equation (14), we
obtain 
\begin{equation}
\left[\begin{array}{c}
\phi_{1}(t)\\
\phi_{2}(t)
\end{array}\right]=\left[\begin{array}{cc}
\cos(\omega t) & -\sin(\omega t)\\
\sin(\omega t) & \cos(\omega t)
\end{array}\right]\left[\begin{array}{c}
\phi_{1}(0)\\
\phi_{2}(0)
\end{array}\right],
\end{equation}
which is simply Eq. (23); the latter is precisely the equation that
gives us the wave function $\Psi(t)$ from $\Psi(0)$ via the evolution
operator. Perhaps, Majorana knew of this simple result, which could
have perfectly well been called a ``Majorana oscillator'' \cite{RefN}.

\subsection{Free Majorana particle in a penetrable box}

To describe a free (one-dimensional) Majorana particle ($\mathcal{S}=0$),
we must first solve the following Dirac equation:
\begin{equation}
\mathrm{i}\hbar\hat{1}_{2}\,\frac{\partial}{\partial t}\Psi=\hat{\mathrm{h}}\Psi=\left(-\mathrm{i}\hbar c\,\hat{\sigma}_{x}\,\frac{\partial}{\partial x}+\mathrm{m}c^{2}\hat{\sigma}_{y}\right)\Psi
\end{equation}

(see Eqs. (9) and (10)). We place the particle inside a
wide interval $\Omega\subset\mathbb{R}$ (e.g., in a wide penetrable
box of size $L$, with ends, for example, at $x=0$ and $x=L$). The
Hamiltonian is an unbounded Hermitian operator, i.e., $\hat{\mathrm{h}}=\hat{\mathrm{h}}^{\dagger}$;
in fact, $\hat{\mathrm{h}}$ is also a self-adjoint operator because
its domain, i.e., the set of Dirac wave functions in $\mathcal{H}=\mathcal{L}^{2}(\Omega)\otimes\mathbb{C}^{2}$
on which $\hat{\mathrm{h}}$ can act ($\equiv\mathcal{D}(\hat{\mathrm{h}})\subset\mathcal{H}$),
includes the periodic boundary condition, $\Psi(L,t)=\Psi(0,t)$ (the
condition that all these functions must satisfy and that we specifically
choose to use in this section), and also $\hat{\mathrm{h}}\Psi\in\mathcal{H}$
(for example, see Ref. \cite{RefB}). Precisely, $\hat{\mathrm{h}}$
satisfies the hermiticity condition (or, in this case, the self-adjointness
condition) 
\begin{equation}
\langle\Phi,\hat{\mathrm{h}}\chi\rangle=\langle\hat{\mathrm{h}}\Phi,\chi\rangle-\left.\mathrm{i}\hbar c\left[\,\Phi^{\dagger}\hat{\sigma}_{x}\,\chi\,\right]\right|_{0}^{L}=\langle\hat{\mathrm{h}}\Phi,\chi\rangle,
\end{equation}
where $\left.\left[\, f\,\right]\right|_{0}^{L}\equiv f(L,t)-f(0,t)$,
and $\Phi$ and $\chi$ are Dirac wave functions in $\mathcal{D}(\hat{\mathrm{h}})=\mathcal{D}(\hat{\mathrm{h}}^{\dagger})$.
Note that the term evaluated at the endpoints of the interval $\Omega$
vanishes because we impose the periodic boundary condition on $\Phi$
and $\chi$. The eigenfunctions of the Hamiltonian are denoted by
$\psi_{\mathrm{p}}^{(+)}$ and $\psi_{\mathrm{p}}^{(-)}$ for the
eigenvalues $E=+E_{\mathrm{p}}=+E_{-\mathrm{p}}=+\sqrt{(c\mathrm{p})^{2}+(\mathrm{m}c^{2})^{2}}\geq\mathrm{m}c^{2}$
(positive-energies) and $E=-E_{\mathrm{p}}\leq-\mathrm{m}c^{2}$ (negative-energies),
respectively:
\begin{equation}
\hat{\mathrm{h}}\,\psi_{\mathrm{p}}^{(+)}=+E_{\mathrm{p}}\,\psi_{\mathrm{p}}^{(+)}\,,\quad\hat{\mathrm{h}}\,\psi_{\mathrm{p}}^{(-)}=-E_{\mathrm{p}}\,\psi_{\mathrm{p}}^{(-)}.
\end{equation}
Explicitly, these eigenfunctions can be written as follows
\begin{equation}
\psi_{\mathrm{p}}^{(+)}(x)=\psi_{\mathrm{p}}^{(+)}\equiv\sqrt{\frac{\mathrm{m}c^{2}}{E_{\mathrm{p}}L}}\, u(\mathrm{p})\exp\left(\mathrm{i}\,\frac{\mathrm{p}}{\hbar}\, x\right)=\sqrt{\frac{1}{2L}}\left[\begin{array}{c}
1\\
\frac{\mathrm{i}\, E_{\mathrm{p}}}{\mathrm{i}\, c\mathrm{p}+\mathrm{m}c^{2}}
\end{array}\right]\exp\left(\mathrm{i}\,\frac{\mathrm{p}}{\hbar}\, x\right)
\end{equation}
and 
\[
\psi_{\mathrm{p}}^{(-)}(x)=\psi_{\mathrm{p}}^{(-)}=(\psi_{-\mathrm{p}}^{(+)})_{C}=(\psi_{-\mathrm{p}}^{(+)})^{*}=\sqrt{\frac{\mathrm{m}c^{2}}{E_{\mathrm{p}}L}}\,(u(-\mathrm{p}))^{*}\exp\left(\mathrm{i}\,\frac{\mathrm{p}}{\hbar}\, x\right)
\]
\begin{equation}
\equiv\sqrt{\frac{\mathrm{m}c^{2}}{E_{\mathrm{p}}L}}\, v(-\mathrm{p})\exp\left(\mathrm{i}\,\frac{\mathrm{p}}{\hbar}\, x\right)=\sqrt{\frac{1}{2L}}\left[\begin{array}{c}
1\\
\frac{-\mathrm{i}\, E_{\mathrm{p}}}{\mathrm{i}\, c\mathrm{p}+\mathrm{m}c^{2}}
\end{array}\right]\exp\left(\mathrm{i}\,\frac{\mathrm{p}}{\hbar}\, x\right),
\end{equation}
where 
\begin{equation}
\mathrm{p}=\hbar\,\frac{2\pi n}{L}\,,\quad n=0,\pm1,\pm2,\ldots,
\end{equation}
because we use the periodic boundary condition (it is understood that
$\mathrm{p}=\mathrm{p}_{n}$, but for simplicity, we do not place
the subscript $n$ on the letter $\mathrm{p}$). Moreover, $\psi_{\mathrm{p}}^{(+)}$
and $\psi_{\mathrm{p}}^{(-)}$ satisfy the following expected relations
of orthonormality and orthogonality:
\begin{equation}
\langle\psi_{\mathrm{p}}^{(+)},\psi_{\mathrm{p'}}^{(+)}\rangle=\langle\psi_{\mathrm{p}}^{(-)},\psi_{\mathrm{p'}}^{(-)}\rangle=\delta_{\mathrm{p}\,\mathrm{p}'}\,,\quad\langle\psi_{\mathrm{p}}^{(+)},\psi_{\mathrm{p'}}^{(-)}\rangle=0,
\end{equation}
where $\delta_{\mathrm{p}\,\mathrm{p}'}$ is the Kronecker delta.
Again, the mutually charge-conjugate vectors, $u(\mathrm{p})$ and
$v(\mathrm{p})$, must satisfy $\mathbb{C}^{2}$-orthogonality and
normalization relations, namely, 
\begin{equation}
(u(\mathrm{p}))^{\dagger}u(\mathrm{p})=(v(-\mathrm{p}))^{\dagger}v(-\mathrm{p})\equiv\frac{E_{\mathrm{p}}}{\mathrm{m}c^{2}}=1\,,\quad(u(\mathrm{p}))^{\dagger}v(-\mathrm{p})=0.
\end{equation}
Thus, the free particle's states $\psi_{\mathrm{p}}^{(+)}(x)$ and
$\psi_{\mathrm{p}}^{(-)}(x)$ are orthonormal eigenstates of the operators
$\hat{\mathrm{p}}$ and $\hat{\mathrm{h}}$, with eigenvalues $\mathrm{p}$
and $E_{\mathrm{p}}$ and $\mathrm{p}$ and $-E_{\mathrm{p}}$, respectively.
However, according to the Dirac hole theory (i.e., abandoning for
a moment Dirac's theory as a single-particle theory), the state $\psi_{\mathrm{\pm\mathrm{p}}}^{(-)}(x)$
can be physically associated with the corresponding antiparticle,
and it would have momentum $\mp\mathrm{p}$ and energy $+E_{\mathrm{p}}$.
Certainly, the operator $\hat{\mathrm{p}}$ for a Dirac particle in
a box is also self-adjoint when the boundary condition $\Psi(L,t)=\Psi(0,t)$
is included in its own domain \cite{RefM,RefO}. The states $\psi_{\mathrm{p}}^{(+)}(x)$
and $\psi_{\mathrm{p}}^{(-)}(x)$ are also eigenstates of $\hat{\mathrm{v}}_{\mathrm{cl}}\equiv c^{2}\,\hat{\mathrm{p}}\,\hat{\mathrm{h}}^{-1}$
with eigenvalues $+c^{2}\mathrm{p}/E_{\mathrm{p}}$ and $-c^{2}\mathrm{p}/E_{\mathrm{p}}$,
respectively. Let us note in passing that, for positive energies,
if $\mathrm{p}\rightarrow\pm\infty$, then $\mathrm{v}_{\mathrm{cl}}\rightarrow\pm c$;
and for negative energies, if $\mathrm{p}\rightarrow\pm\infty$, then
$\mathrm{v}_{\mathrm{cl}}\rightarrow\mp c$, i.e., the spectrum of
$\hat{\mathrm{v}}_{\mathrm{cl}}$ is the interval $[-c,+c]$. Thus,
this operator is certainly bounded and self-adjoint \cite{RefL}.
Notice finally that the periodic boundary condition did not provide
non-trivial eigensolutions for the range of energies $-\mathrm{m}c^{2}<E<+\mathrm{m}c^{2}$.

The most general wave function at $t=0$ can be written as
\[
\Psi(x,0)=\sum_{\mathrm{p=-\infty}}^{+\infty}\left[\, c_{\mathrm{p}}^{(+)}(0)\,\psi_{\mathrm{p}}^{(+)}(x)+c_{\mathrm{p}}^{(-)}(0)\,\psi_{\mathrm{p}}^{(-)}(x)\,\right]
\]
\begin{equation}
=\sum_{\mathrm{p=-\infty}}^{+\infty}\left[\, c_{\mathrm{p}}^{(+)}(0)\,\psi_{\mathrm{p}}^{(+)}(x)+c_{\mathrm{-\mathrm{p}}}^{(-)}(0)\,\psi_{\mathrm{-\mathrm{p}}}^{(-)}(x)\,\right],
\end{equation}
where in the last expression we used the fact that $\sum_{\mathrm{p}}f_{\mathrm{p}}=\sum_{\mathrm{p}}f_{-\mathrm{p}}$.
In addition, $c_{\mathrm{p}}^{(\pm)}(0)\equiv c_{\mathrm{p}}^{(\pm)}(t=0)$
are arbitrary complex constants. The latter quantities satisfy a constraint,
namely,
\begin{equation}
\sum_{\mathrm{p=-\infty}}^{+\infty}\left[\,\left|\, c_{\mathrm{p}}^{(+)}(0)\,\right|^{2}+\left|\, c_{\mathrm{p}}^{(-)}(0)\,\right|^{2}\,\right]=\sum_{\mathrm{p=-\infty}}^{+\infty}\left[\,\left|\, c_{\mathrm{p}}^{(+)}(0)\,\right|^{2}+\left|\, c_{\mathrm{\mathrm{-p}}}^{(-)}(0)\,\right|^{2}\,\right]=1,
\end{equation}
provided that $\Psi(x,0)$ is a normalized wave function. Again, in
a hole-theoretic description, the destruction (or removal from the
``vacuum'') of a negative-energy particle of momentum $-\mathrm{p}$
appears as the creation of its corresponding positive-energy antiparticle
with momentum $+\mathrm{p}$ (a hole in the ``vacuum''). We also
write the sum in Eq. (39) in terms of $\psi_{\mathrm{-\mathrm{p}}}^{(-)}(x)$
just to highlight this point. In this way, we can interpret $c_{\mathrm{\mathrm{-\mathrm{p}}}}^{(-)}(0)$
($\, c_{\mathrm{p}}^{(+)}(0)\,$) as the amplitude for an antiparticle
(a particle) state with momentum $+\mathrm{p}$ and energy $+E_{\mathrm{p}}$.
These complex-valued objects could be redefined as $c_{\mathrm{p}}^{(+)}(0)\equiv b_{\mathrm{p}}$
and $c_{\mathrm{\mathrm{-\mathrm{p}}}}^{(-)}(0)\equiv d_{\mathrm{p}}^{*}$,
and in the formalism of second quantization, they become time-independent
operators, namely, $b_{\mathrm{p}}\rightarrow\hat{b}_{\mathrm{p}}$
(the annihilation or absorption operator for the particle) and $d_{\mathrm{p}}^{*}\rightarrow\hat{d}_{\mathrm{p}}^{\dagger}$
(the creation operator for the antiparticle). In particle physics,
it is customary to use these symbols.

Once again, because we are describing a Majorana particle, the initial
state must obey the Majorana condition, $\Psi(x,0)=\Psi_{C}(x,0)=(\Psi(x,0))^{*}$.
With the latter condition imposed upon $\Psi(x,0)$ (Eq. (39)), we
obtain the result $c_{\mathrm{-p}}^{(-)}(0)=(c_{\mathrm{p}}^{(+)}(0))^{*}$,
i.e., $d_{\mathrm{p}}^{*}=(b_{\mathrm{p}})^{*}\Rightarrow b_{\mathrm{p}}=d_{\mathrm{p}}$.
Therefore, the most general initial wave function for the free Majorana
particle in this penetrable box takes the form
\begin{equation}
\Psi(x,0)=\sum_{\mathrm{p=-\infty}}^{+\infty}\left[\, c_{\mathrm{p}}^{(+)}(0)\,\psi_{\mathrm{p}}^{(+)}(x)+\mathrm{c.c.}\,\right]\,,\quad\sum_{\mathrm{p=-\infty}}^{+\infty}\left|\, c_{\mathrm{p}}^{(+)}(0)\,\right|^{2}=\frac{1}{2}.
\end{equation}
It is then clear that  here we need only one type of constant, $b_{\mathrm{p}}$
or $d_{\mathrm{p}}$. Abandoning the language of the single-particle
theory itself, we could say that a Majorana particle is a superposition
of a particle and its antiparticle, but this particle and antiparticle
coincide, as expected.

The wave function $\Psi(x,t)$ can be obtained by applying the evolution
operator $\hat{U}(t)=\exp\left(-\frac{\mathrm{i}}{\hbar}\, t\,\hat{\mathrm{h}}\right)$
to the initial wave function $\Psi(x,0)$, but only formally. In fact,
in this case, we cannot use the power series expansion of the exponential,
because $\hat{\mathrm{h}}$ is an unbounded operator, i.e., it cannot
act on all the wave functions of the Hilbert space of the system.
However, the correct result is the same as that obtained following
the usual prescription (for example, see Ref. \cite{RefP}), namely,
\begin{equation}
\Psi(x,t)=\sum_{\mathrm{p=-\infty}}^{+\infty}\left[\, c_{\mathrm{p}}^{(+)}(0)\,\psi_{\mathrm{p}}^{(+)}(x)\exp\left(-\mathrm{i}\,\frac{E_{\mathrm{p}}}{\hbar}\, t\right)+\mathrm{c.c.}\,\right],
\end{equation}
with $\sum_{\mathrm{p=-\infty}}^{+\infty}\left|\, c_{\mathrm{p}}^{(+)}(0)\,\right|^{2}=\frac{1}{2}$.
Notice again that $\hat{U}(t)$ is a real operator and therefore preserves
the real character of the initial wave function. Additionally, notice
that we would have to multiply the c.c. term in Eqs. (41) and (42)
by $\exp(\mathrm{i}\theta)$ in the case of using the generalized
Majorana condition. In the end, the real wave function in Eq. (42)
can also be written as follows: 
\begin{equation}
\Psi(x,t)=\sqrt{\frac{1}{L}}\sum_{\mathrm{p=-\infty}}^{+\infty}\sqrt{\frac{\mathrm{m}c^{2}}{E_{\mathrm{p}}}}\left[\, b_{\mathrm{p}}\, u(\mathrm{p})\exp\left(\mathrm{i}\,\frac{\mathrm{p}}{\hbar}\, x-\mathrm{i}\,\frac{E_{\mathrm{p}}}{\hbar}\, t\right)+\mathrm{c.c.}\,\right],
\end{equation}
where $\sum_{\mathrm{p=-\infty}}^{+\infty}\left|\, b_{\mathrm{p}}\,\right|^{2}=\frac{1}{2}$.
In the second quantized theory, the corresponding Majorana quantum
field operator is Hermitian, i.e., $\Psi(x,t)\rightarrow\hat{\Psi}(x,t)=\hat{\Psi}^{\dagger}(x,t)$.
This is because the wave function $\Psi(x,t)$, i.e., the classical
field, is real. 

It is worth mentioning that the boundary condition we have used in
this section, i.e., the periodic boundary condition, is just one of
the infinite non-confining boundary conditions that a (one-dimensional)
Majorana particle can support when it is inside a box, namely,
\begin{equation}
\Psi(L,t)=-\frac{(\mathrm{i}m_{0}\,\hat{\sigma}_{y}+\hat{\sigma}_{x})}{m_{2}}\,\Psi(0,t),
\end{equation}
with $m_{2}\neq0$ (and thus Eq. (44) is a non-confining family of
boundary conditions) and $(m_{0})^{2}+(m_{2})^{2}=1$. In addition,
\begin{equation}
\Psi(L,t)=\frac{m_{3}\,\hat{\sigma}_{z}+\hat{1}_{2}}{m_{1}}\,\Psi(0,t),
\end{equation}
with $m_{1}\neq0$ (and thus Eq. (45) is a non-confining family of
boundary conditions) and $(m_{1})^{2}+(m_{3})^{2}=1$ (see Eqs. (67)
and (68) in Ref. \cite{RefB}). Specifically, the periodic boundary
condition, $\Psi(L,t)=\Psi(0,t)$, can be obtained from Eq. (45) by
making $m_{1}=1$ and $m_{3}=0$. Certainly, none of the boundary
conditions in Eqs. (44) and (45) cancels the probability current density
at the ends of the box (which is why we call them non-confining boundary
conditions) \cite{RefB}. Naturally, all these boundary conditions
cancel the boundary term in Eq. (32) (this is because when all of
them are included in $\mathcal{D}(\hat{\mathrm{h}})$, the relation
$\mathcal{D}(\hat{\mathrm{h}})=\mathcal{D}(\hat{\mathrm{h}}^{\dagger})$
with $\hat{\mathrm{h}}=\hat{\mathrm{h}}^{\dagger}$ is satisfied)
\cite{RefB}. Therefore, if we impose $\Phi=\chi=\Psi$ in Eq. (32),
we obtain, as before, the expression $\langle\Psi,\hat{\mathrm{h}}\Psi\rangle\equiv\langle\hat{\mathrm{h}}\rangle_{\Psi}=\langle\hat{\mathrm{h}}\Psi,\Psi\rangle=\langle\Psi,\hat{\mathrm{h}}\Psi\rangle^{*}=\langle\hat{\mathrm{h}}\rangle_{\Psi}^{*}$.
The latter result together with Eq. (26) (which is verified whenever
$\Psi$ is real-valued) leads to the result given in Eq. (28), i.e.,
$\langle\hat{\mathrm{h}}\rangle_{\Psi}=0$. Clearly, the boundary
conditions in Eqs. (44) and (45) are real boundary conditions when
the wave function describes a Majorana particle, i.e., when the wave
function is real-valued. Lastly, the results obtained in this subsection
obviously apply to those of subsection A if we only maintain the eigenvalue
$\mathrm{p}=0$. 

\subsection{Free Majorana particle in an impenetrable box}

To describe a free (one-dimensional) Majorana particle ($\mathcal{S}=0$)
that is inside an impenetrable box of width $L$, we must first solve
Eq. (31). In this case, the self-adjoint (Dirac) Hamiltonian operator
has a domain that includes a general two-real-parameter family of
confining boundary conditions, i.e., a two-real-parameter family at
one end of the box, and another family with the same two parameters
at the other end \cite{RefQ}. These boundary conditions cancel the
probability current density at the ends of the box (which is why we
call them confining boundary conditions) \cite{RefB,RefQ}. Likewise,
these boundary conditions imposed upon $\Phi$ and $\chi$ in Eq.
(32) lead to the vanishing of the boundary term present there (this
is simply because $\hat{\mathrm{h}}$, with these boundary conditions
within its domain, is a self-adjoint operator). As it was demonstrated
in Ref. \cite{RefB}, from all the boundary conditions that confine
a (one-dimensional) Dirac particle inside a box, only four of them
can be used to confine a (one-dimensional) Majorana particle (due
to the Majorana condition), namely, 
\begin{equation}
\phi_{2}(L,t)=\phi_{2}(0,t)=0,
\end{equation}
and 
\begin{equation}
\phi_{1}(L,t)=\phi_{1}(0,t)=0.
\end{equation}
These two boundary conditions are the Dirichlet boundary condition
imposed upon the lower component (Eq. (46)) and the upper component
of the wave function (Eq. (47)). Likewise, 
\begin{equation}
\phi_{2}(L,t)=\phi_{1}(0,t)=0,
\end{equation}
and 
\begin{equation}
\phi_{1}(L,t)=\phi_{2}(0,t)=0.
\end{equation}
These are two mixed boundary conditions. The authors of Ref. \cite{RefK}
also obtained the result that there are just four confining boundary
conditions for a (one-dimensional) Majorana particle in a box (see
Eq. (3.8) therein), but they expressed these boundary conditions only
in terms of a single component of the Dirac wave function in the Dirac
representation. Note that, by taking the limit $m_{2}\rightarrow0$
in Eq. (44) and its respective inverse, we obtain the results in Eq.
(46) (when $m_{0}=1$) and Eq. (47) (when $m_{0}=-1$). Likewise,
by taking the limit $m_{1}\rightarrow0$ in Eq. (45) and its respective
inverse, we obtain the results in Eq. (48) (when $m_{3}=1$) and Eq.
(49) (when $m_{3}=-1$) \cite{RefB}. Thus, the two one-parameter
families of boundary conditions in Eqs. (44) and (45) can also generate
the four confining boundary conditions. Consequently, these two one-parameter
families really make up the most general set of boundary conditions
for the (one-dimensional) Majorana particle in a box. Unexpectedly,
the four confining boundary conditions for the (one-dimensional) Majorana
particle are exactly the four boundary conditions that mathematically
can arise from the general linear boundary condition used in the MIT
bag model for a hadronic structure (certainly, the four boundary conditions
in this model must also obey the Majorana condition) \cite{RefB},
and the boundary condition in Eq. (49) is precisely the boundary condition
commonly used in that model, but in one dimension \cite{RefB}.

Once again, we use the following notation to identify the normalized
eigenfunctions of the Hamiltonian in Eq. (31): $\psi_{\mathrm{p}}^{(+)}$
and $\psi_{\mathrm{p}}^{(-)}$, for the eigenvalues $E=+E_{\mathrm{p}}=+E_{-\mathrm{p}}=+\sqrt{(c\mathrm{p})^{2}+(\mathrm{m}c^{2})^{2}}\geq\mathrm{m}c^{2}$
(positive-energies) and $E=-E_{\mathrm{p}}\leq-\mathrm{m}c^{2}$ (negative-energies),
respectively (see Eq, (33)). Likewise, we write $\psi_{\mathrm{q}}^{(+)}$
and $\psi_{\mathrm{q}}^{(-)}$ for the eigenvalues $E=+E_{\mathrm{q}}=+E_{-\mathrm{q}}=+\sqrt{(\mathrm{m}c^{2})^{2}-(c\mathrm{q})^{2}}<\mathrm{m}c^{2}$
(positive-energies) and $E=-E_{\mathrm{q}}>-\mathrm{m}c^{2}$ (negative-energies),
respectively. The latter eigenfunctions can be obtained from the former
with the replacement $\mathrm{p}\rightarrow-\mathrm{i}\mathrm{q}$,
where $\mathrm{q}>0$; thus, $\psi_{\mathrm{p}}^{(\pm)}\rightarrow\psi_{\mathrm{q}}^{(\pm)}$,
and $E_{\mathrm{p}}\rightarrow E_{\mathrm{q}}$. 

Naturally, for each of the four confining boundary conditions, we
will have a different orthonormal system of eigenstates of $\hat{\mathrm{h}}$.
However, all these eigenstates will not be eigenstates of the Dirac
momentum operator $\hat{\mathrm{p}}=-\mathrm{i}\hbar\hat{1}_{2}\,\partial/\partial x$;
thus, the letter $\mathrm{p}$ in the eigenfunctions will just be
a label. Moreover, the operator $\hat{\mathrm{p}}$ in an interval
is self-adjoint just when $\Psi(L,t)=\hat{M}\Psi(0,t)$, where $\hat{M}$
is a unitary matrix, but the confining boundary conditions in Eqs.
(46)-(49) are not included in that latter general boundary condition
\cite{RefM,RefO}. Thus, in the present problem, we do not have a
completely (physically) acceptable momentum operator. However, if
we could instantly remove the walls of the box and immediately measure
the momentum, that would put the particle in an eigenstate of the
momentum operator, which would be, in this case, a self-adjoint operator
with a continuous spectrum (see, for example, p. 269 in Ref. \cite{RefR}).
Explicitly, the eigenfunctions of the Hamiltonian can be written as
follows. 

For $\phi_{2}(L,t)=\phi_{2}(0,t)=0$, 
\begin{equation}
\psi_{\mathrm{p}}^{(+)}(x)=\psi_{\mathrm{p}}^{(+)}=\sqrt{\frac{1}{L}}\left[\begin{array}{c}
\frac{c\mathrm{p}}{E_{\mathrm{p}}}\cos\left(\frac{\mathrm{p}x}{\hbar}\right)+\frac{\mathrm{m}c^{2}}{E_{\mathrm{p}}}\sin\left(\frac{\mathrm{p}x}{\hbar}\right)\\
\mathrm{i}\sin\left(\frac{\mathrm{p}x}{\hbar}\right)
\end{array}\right]
\end{equation}
and 
\begin{equation}
\psi_{\mathrm{p}}^{(-)}(x)=\psi_{\mathrm{p}}^{(-)}=(\psi_{-\mathrm{p}}^{(+)})_{C}=(\psi_{-\mathrm{p}}^{(+)})^{*}=\sqrt{\frac{1}{L}}\left[\begin{array}{c}
-\frac{c\mathrm{p}}{E_{\mathrm{p}}}\cos\left(\frac{\mathrm{p}x}{\hbar}\right)-\frac{\mathrm{m}c^{2}}{E_{\mathrm{p}}}\sin\left(\frac{\mathrm{p}x}{\hbar}\right)\\
\mathrm{i}\sin\left(\frac{\mathrm{p}x}{\hbar}\right)
\end{array}\right]
\end{equation}
where 
\begin{equation}
\mathrm{p}=\hbar\,\frac{\pi N}{L}\,,\quad N=1,2,3,\ldots\,.
\end{equation}

For $\phi_{1}(L,t)=\phi_{1}(0,t)=0$,
\begin{equation}
\psi_{\mathrm{p}}^{(+)}(x)=\psi_{\mathrm{p}}^{(+)}=\sqrt{\frac{1}{L}}\left[\begin{array}{c}
\mathrm{i}\sin\left(\frac{\mathrm{p}x}{\hbar}\right)\\
\frac{c\mathrm{p}}{E_{\mathrm{p}}}\cos\left(\frac{\mathrm{p}x}{\hbar}\right)-\frac{\mathrm{m}c^{2}}{E_{\mathrm{p}}}\sin\left(\frac{\mathrm{p}x}{\hbar}\right)
\end{array}\right],
\end{equation}
and 
\begin{equation}
\psi_{\mathrm{p}}^{(-)}(x)=\psi_{\mathrm{p}}^{(-)}=(\psi_{-\mathrm{p}}^{(+)})_{C}=(\psi_{-\mathrm{p}}^{(+)})^{*}=\sqrt{\frac{1}{L}}\left[\begin{array}{c}
\mathrm{i}\sin\left(\frac{\mathrm{p}x}{\hbar}\right)\\
-\frac{c\mathrm{p}}{E_{\mathrm{p}}}\cos\left(\frac{\mathrm{p}x}{\hbar}\right)+\frac{\mathrm{m}c^{2}}{E_{\mathrm{p}}}\sin\left(\frac{\mathrm{p}x}{\hbar}\right)
\end{array}\right],
\end{equation}
where
\begin{equation}
\mathrm{p}=\hbar\,\frac{\pi N}{L}\,,\quad N=1,2,3,\ldots\,.
\end{equation}
Notice that these two boundary conditions do not provide non-trivial
eigensolutions in the range of energies $-\mathrm{m}c^{2}\leq E\leq+\mathrm{m}c^{2}$.

For $\phi_{2}(L,t)=\phi_{1}(0,t)=0$, in this case, the eigenfunctions
of the Hamiltonian for the energies $E>\mathrm{m}c^{2}$ and $E<-\mathrm{m}c^{2}$
are precisely $\psi_{\mathrm{p}}^{(+)}\mbox{ and }\psi_{\mathrm{p}}^{(-)}$,
respectively, given by Eqs. (53) and (54). However, the spectrum of
$\hat{\mathrm{h}}$ is obtained from the following transcendental
equation
\begin{equation}
\frac{\hbar}{\mathrm{m}c}\frac{1}{L}\frac{\mathrm{p}L}{\hbar}=\tan\left(\frac{\mathrm{p}L}{\hbar}\right),
\end{equation}
whose solutions must obey the inequality $\mathrm{p}>0$. The eigensolution
for $\mathrm{p}=0$ ($\Rightarrow E=\pm\mathrm{m}c^{2}$) is the trivial
solution whenever $L\neq\hbar/\mathrm{m}c$. Moreover, we have one
positive-energy eigenfunction,
\begin{equation}
\psi_{\mathrm{q}}^{(+)}(x)=\psi_{\mathrm{q}}^{(+)}=\sqrt{\frac{1}{L}}\left[\begin{array}{c}
\sinh\left(\frac{\mathrm{q}x}{\hbar}\right)\\
-\frac{\mathrm{i}\, c\mathrm{q}}{E_{\mathrm{q}}}\cosh\left(\frac{\mathrm{q}x}{\hbar}\right)+\frac{\mathrm{i}\,\mathrm{m}c^{2}}{E_{\mathrm{q}}}\sinh\left(\frac{\mathrm{q}x}{\hbar}\right)
\end{array}\right],
\end{equation}
with energy $E=+E_{\mathrm{q}}$, and the corresponding negative-energy
eigensolution,
\begin{equation}
\psi_{\mathrm{q}}^{(-)}(x)=\psi_{\mathrm{q}}^{(-)}=(\psi_{-\mathrm{q}}^{(+)})_{C}=(\psi_{-\mathrm{q}}^{(+)})^{*}=(-1)\sqrt{\frac{1}{L}}\left[\begin{array}{c}
\sinh\left(\frac{\mathrm{q}x}{\hbar}\right)\\
\frac{\mathrm{i}\, c\mathrm{q}}{E_{\mathrm{q}}}\cosh\left(\frac{\mathrm{q}x}{\hbar}\right)-\frac{\mathrm{i}\,\mathrm{m}c^{2}}{E_{\mathrm{q}}}\sinh\left(\frac{\mathrm{q}x}{\hbar}\right)
\end{array}\right],
\end{equation}
with energy $E=-E_{\mathrm{q}}$. The quantity $\mathrm{q}$ is the
only positive solution of the equation 
\begin{equation}
\frac{\hbar}{\mathrm{m}c}\frac{1}{L}\frac{\mathrm{q}L}{\hbar}=\tanh\left(\frac{\mathrm{q}L}{\hbar}\right),
\end{equation}
where $L>\hbar/\mathrm{m}c$. 

For $\phi_{1}(L,t)=\phi_{2}(0,t)=0$, in this case, the eigenfunctions
of the Hamiltonian are precisely $\psi_{\mathrm{p}}^{(+)}\mbox{ and }\psi_{\mathrm{p}}^{(-)}$
given by Eqs. (50) and (51). Again, the spectrum of $\hat{\mathrm{h}}$
is obtained from a transcendental equation, namely,
\begin{equation}
-\frac{\hbar}{\mathrm{m}c}\frac{1}{L}\frac{\mathrm{p}L}{\hbar}=\tan\left(\frac{\mathrm{p}L}{\hbar}\right),
\end{equation}
and its solutions must satisfy the inequality $\mathrm{p}>0$. In
this case, the eigensolution for $\mathrm{p}=0$ ($\Rightarrow E=\pm\mathrm{m}c^{2}$)
is just the trivial solution. Likewise, we do not have non-trivial
eigensolutions in the range of energies $-\mathrm{m}c^{2}<E<+\mathrm{m}c^{2}$. 

As expected, the eigenvectors corresponding to each boundary condition
satisfy the relations of orthonormality and orthogonality identical
to those given in Eq. (37). Additionally, the states (57) and (58)
verify the following relations 
\begin{equation}
\langle\psi_{\mathrm{q}}^{(+)},\psi_{\mathrm{q}}^{(+)}\rangle=\langle\psi_{\mathrm{q}}^{(-)},\psi_{\mathrm{q}}^{(-)}\rangle=1\,,\quad\langle\psi_{\mathrm{q}}^{(+)},\psi_{\mathrm{q}}^{(-)}\rangle=0,
\end{equation}
and also
\begin{equation}
\langle\psi_{\mathrm{q}}^{(+)},\psi_{\mathrm{p}}^{(+)}\rangle=\langle\psi_{\mathrm{q}}^{(+)},\psi_{\mathrm{p}}^{(-)}\rangle=\langle\psi_{\mathrm{q}}^{(-)},\psi_{\mathrm{p}}^{(+)}\rangle=\langle\psi_{\mathrm{q}}^{(-)},\psi_{\mathrm{p}}^{(-)}\rangle=0,
\end{equation}
for all $\mathrm{p}$ that is solution of Eq. (56).

Again, the most general normalized Dirac wave function at $t=0$ can
be written similarly to that in Eq. (39) and is subject to a restriction
similar to that in Eq. (40), namely, 
\begin{equation}
\Psi(x,0)=c_{\mathrm{q}}^{(+)}(0)\,\psi_{\mathrm{q}}^{(+)}(x)+c_{\mathrm{q}}^{(-)}(0)\,\psi_{\mathrm{q}}^{(-)}(x)+\sum_{\mathrm{p}}^{+\infty}\left[\, c_{\mathrm{p}}^{(+)}(0)\,\psi_{\mathrm{p}}^{(+)}(x)+c_{\mathrm{p}}^{(-)}(0)\,\psi_{\mathrm{p}}^{(-)}(x)\,\right],
\end{equation}
and
\begin{equation}
\left|\, c_{\mathrm{q}}^{(+)}(0)\,\right|^{2}+\left|\, c_{\mathrm{q}}^{(-)}(0)\,\right|^{2}+\sum_{\mathrm{p}}^{+\infty}\left[\,\left|\, c_{\mathrm{p}}^{(+)}(0)\,\right|^{2}+\left|\, c_{\mathrm{p}}^{(-)}(0)\,\right|^{2}\,\right]=1,
\end{equation}
respectively (also, $c_{\mathrm{q}}^{(\pm)}(0)\equiv c_{\mathrm{q}}^{(\pm)}(t=0)$
and $c_{\mathrm{p}}^{(\pm)}(0)\equiv c_{\mathrm{p}}^{(\pm)}(t=0)$,
as before). Certainly, in each case, the infinite sum is determined
by the relations given in Eqs. (52), (55), (56) and (59). Moreover,
when considering the boundary conditions (46), (47) and (49), only
the terms with sums should be written in Eqs. (63) and (64), i.e.,
the terms depending on $\mathrm{q}$ are not present in those cases.
The initial state for a Majorana particle must satisfy the Majorana
condition, $\Psi(x,0)=\Psi_{C}(x,0)=(\Psi(x,0))^{*}$. The latter
condition together with the relations $(\psi_{\mathrm{p}}^{(\pm)})^{*}=\psi_{-\mathrm{p}}^{(\mp)}=-\psi_{\mathrm{p}}^{(\mp)}$
and $(\psi_{\mathrm{q}}^{(\pm)})^{*}=\psi_{-\mathrm{q}}^{(\mp)}=-\psi_{\mathrm{q}}^{(\mp)}$
(in the last step of each of these equalities, we use the fact that
all these eigenfunctions are odd in $\mathrm{p}$ and $\mathrm{q}$)
leads to the relations $c_{\mathrm{p}}^{(-)}(0)=-(c_{\mathrm{p}}^{(+)}(0))^{*}$
and $c_{\mathrm{q}}^{(-)}(0)=-(c_{\mathrm{q}}^{(+)}(0))^{*}$. Substituting
the latter relations in Eq. (63), as well as those given before for
the eigenfunctions, we can write 
\begin{equation}
\Psi(x,0)=\left[\, c_{\mathrm{q}}^{(+)}(0)\,\psi_{\mathrm{q}}^{(+)}(x)+\mathrm{c.c.}\,\right]+\sum_{\mathrm{p}}^{+\infty}\left[\, c_{\mathrm{p}}^{(+)}(0)\,\psi_{\mathrm{p}}^{(+)}(x)+\mathrm{c.c.}\,\right],
\end{equation}
with $\left|\, c_{\mathrm{q}}^{(+)}(0)\,\right|^{2}+\sum_{\mathrm{p}}^{+\infty}\left|\, c_{\mathrm{p}}^{(+)}(0)\,\right|^{2}=\frac{1}{2}$. 

Again, the real general solution $\Psi(x,t)$ is simply given by
\[
\Psi(x,t)=\left[\, c_{\mathrm{q}}^{(+)}(0)\,\psi_{\mathrm{q}}^{(+)}(x)\exp\left(-\mathrm{i}\,\frac{E_{\mathrm{q}}}{\hbar}\, t\right)+\mathrm{c.c.}\,\right]
\]
\begin{equation}
+\sum_{\mathrm{p}}^{+\infty}\left[\, c_{\mathrm{p}}^{(+)}(0)\,\psi_{\mathrm{p}}^{(+)}(x)\exp\left(-\mathrm{i}\,\frac{E_{\mathrm{p}}}{\hbar}\, t\right)+\mathrm{c.c.}\,\right],
\end{equation}
with the restriction that follows Eq. (65), which comes from the normalization
of the initial state and that always remains equal to one. The correct
result given by Eq. (66) can be obtained by formally applying the
evolution operator to the wave function $\Psi(x,0)$. In fact, the
four Hamiltonian operators considered here act in the manner presented
in Eq. (31) but do not have equal domains because each of these has
a different boundary condition. Certainly, they are unbounded operators,
by acting only on a subset of the Hilbert space. 

\subsection{A Majorana particle in a linear Lorentz scalar potential}

In this case, we choose $\mathcal{S}=kx$, where $k>0$
is a constant. The Dirac equation we must solve is
\begin{equation}
\mathrm{i}\hbar\hat{1}_{2}\,\frac{\partial}{\partial t}\Psi=\hat{\mathrm{h}}\Psi=\left[\,-\mathrm{i}\hbar c\,\hat{\sigma}_{x}\,\frac{\partial}{\partial x}+(kx+\mathrm{m}c^{2})\hat{\sigma}_{y}\,\right]\Psi
\end{equation}
(see Eqs. (9) and (10)). We have a particle inside the infinite region
$\Omega=\mathbb{R}$, i.e., the entire real line. In general, the
Hamiltonian operator in the Hilbert space $\mathcal{H}=\mathcal{L}^{2}(\Omega)\otimes\mathbb{C}^{2}$
is essentially self-adjoint on the set of infinitely differentiable
two-component functions with compact support in $\Omega$, no matter
how great the potential $\mathcal{S}$ is at infinity \cite{RefS}.
In fact, $\hat{\mathrm{h}}$ is a Hermitian operator on the same set
of functions, i.e., it satisfies the hermiticity condition in Eq.
(32) with $0\rightarrow-\infty$ and $L\rightarrow+\infty$ (this
is because all these functions vanish at infinity), but $\hat{\mathrm{h}}$
is self-adjoint on the Sobolev space $W^{1,2}(\mathbb{R})^{2}$. We
denote the eigenfunctions of the Hamiltonian by $\psi_{0}$, $\psi_{N}^{(+)}$,
and $\psi_{N}^{(-)}$, for the eigenvalues $E=0$, $E=+\epsilon_{N}=+\sqrt{2\hbar ckN}$
(positive-energies), and $E=-\epsilon_{N}$ (negative-energies), with
$N=1,2,\ldots$, respectively:
\begin{equation}
\hat{\mathrm{h}}\,\psi_{0}=0\psi_{0}=0\,,\quad\hat{\mathrm{h}}\,\psi_{N}^{(+)}=+\epsilon_{N}\,\psi_{N}^{(+)}\,,\quad\hat{\mathrm{h}}\,\psi_{N}^{(-)}=-\epsilon_{N}\,\psi_{N}^{(-)}.
\end{equation}
Specifically, it was shown in Ref. \cite{RefT} that the spectrum
of $\hat{\mathrm{h}}$ is purely discrete and unbounded above and
below. Explicitly, the eigenfunctions can be written as follows
\begin{equation}
\psi_{0}(x)=\psi_{0}=a_{0}\left[\begin{array}{c}
0\\
1
\end{array}\right]\exp\left(-\frac{k}{\hbar c}\frac{(x+x_{0})^{2}}{2}\right),
\end{equation}
and
\begin{equation}
\psi_{N}^{(+)}(x)=\psi_{N}^{(+)}=a_{N}\left[\begin{array}{c}
-\mathrm{i}\sqrt{2N}\, H_{N-1}\left(\sqrt{\frac{k}{\hbar c}}(x+x_{0})\right)\\
H_{N}\left(\sqrt{\frac{k}{\hbar c}}(x+x_{0})\right)
\end{array}\right]\exp\left(-\frac{k}{\hbar c}\frac{(x+x_{0})^{2}}{2}\right),
\end{equation}
and
\begin{equation}
\psi_{N}^{(-)}(x)=\psi_{N}^{(-)}=\hat{\sigma}_{z}\psi_{N}^{(+)}=a_{N}\left[\begin{array}{c}
-\mathrm{i}\sqrt{2N}\, H_{N-1}\left(\sqrt{\frac{k}{\hbar c}}(x+x_{0})\right)\\
-H_{N}\left(\sqrt{\frac{k}{\hbar c}}(x+x_{0})\right)
\end{array}\right]\exp\left(-\frac{k}{\hbar c}\frac{(x+x_{0})^{2}}{2}\right),
\end{equation}
where $N=1,2,\ldots$. In Eqs. (69)-(71), $a_{0}$ and $a_{N}$ are
(real) normalization constants, $x_{0}\equiv\mathrm{m}c^{2}/k$, and
the functions $H_{N}$ are Hermite polynomials. Additionally, notice
that the complex eigenfunction $\psi_{N}^{(-)}$ is the same physical
eigenstate as $(\psi_{N}^{(+)})^{*}=(\psi_{N}^{(+)})_{C}$, as expected.
As usual, these eigenvectors satisfy the relations of orthogonality,
namely, 
\begin{equation}
\langle\psi_{N}^{(\pm)},\psi_{N'}^{(\pm)}\rangle=0\,,\:\mathrm{with}\,\: N\neq N'\,,\quad\langle\psi_{0},\psi_{N}^{(\pm)}\rangle=0\,,\quad\langle\psi_{N}^{(+)},\psi_{N'}^{(-)}\rangle=0.
\end{equation}

The most general Dirac wave function at $t=0$ can be written as
\begin{equation}
\Psi(x,0)=\sum_{N=1}^{+\infty}c_{N}^{(+)}(0)\,\psi_{N}^{(+)}(x)+c_{0}(0)\,\psi_{0}(x)+\sum_{N=1}^{+\infty}c_{N}^{(-)}(0)\,\psi_{N}^{(-)}(x),
\end{equation}
and assuming that the eigenfunctions are normalized, the condition
$\left\Vert \,\Psi(x,0)\,\right\Vert =1$ leads to the restriction
\begin{equation}
\sum_{N=1}^{+\infty}\left|\, c_{N}^{(+)}(0)\,\right|^{2}+\left|\, c_{0}(0)\,\right|^{2}+\sum_{N=1}^{+\infty}\left|\, c_{N}^{(-)}(0)\,\right|^{2}=1,
\end{equation}
where $c_{N}^{(\pm)}(0)\equiv c_{N}^{(\pm)}(t=0)$ and $c_{0}(0)\equiv c_{0}(t=0)$
. The initial state for a Majorana particle is obtained by imposing
the Majorana condition ($\Psi(x,0)=\Psi_{C}(x,0)=(\Psi(x,0))^{*}$)
in Eq. (73), and this leads to the relations $c_{N}^{(-)}(0)=(c_{N}^{(+)}(0))^{*}$
and $c_{0}(0)=(c_{0}(0))^{*}$. Substituting the latter relations
in Eq. (73), as well as that given before for the eigenfunctions,
we can write
\begin{equation}
\Psi(x,0)=c_{0}(0)\,\psi_{0}(x)+\sum_{N=1}^{+\infty}\left[\, c_{N}^{(+)}(0)\,\psi_{N}^{(+)}(x)+\mathrm{c.c.}\,\right],
\end{equation}
with $\frac{1}{2}\left|\, c_{0}(0)\,\right|^{2}+\sum_{N=1}^{+\infty}\left|\, c_{N}^{(+)}(0)\,\right|^{2}=\frac{1}{2}$.

Once again, the real general solution $\Psi(x,t)$ is simply given
by
\begin{equation}
\Psi(x,t)=c_{0}(0)\,\psi_{0}(x)+\sum_{N=1}^{+\infty}\left[\, c_{N}^{(+)}(0)\,\psi_{N}^{(+)}(x)\exp\left(-\mathrm{i}\,\frac{\epsilon_{N}}{\hbar}\, t\right)+\mathrm{c.c.}\,\right],
\end{equation}
with the restriction that follows Eq. (75). Recently, for the problem
addressed in this subsection, real particular solutions for a given
$N$ were obtained using the factorization method or the so-called
supersymmetric procedure \cite{RefU}. The gamma matrices used therein
defined another Majorana representation, namely, $\hat{\gamma}^{0}\,''=-\hat{\sigma}_{y}$
and $\hat{\gamma}^{1}\,''=+\mathrm{i}\hat{\sigma}_{z}$. These matrices
and the matrices used in the present article ($\hat{\gamma}^{0}=+\hat{\sigma}_{y}$
and $\hat{\gamma}^{1}=-\mathrm{i}\hat{\sigma}_{z}$) are related by
$\hat{\gamma}^{\mu}\,''=\hat{\sigma}_{x}\,\hat{\gamma}^{\mu}\,\hat{\sigma}_{x}$
and $\left[\,\phi_{1}''\;\phi_{2}''\,\right]^{\mathrm{T}}=\hat{\sigma}_{x}\left[\,\phi_{1}\;\phi_{2}\,\right]^{\mathrm{T}}$. 

\section{Conclusions}

A Majorana particle is its own antiparticle. Therefore,
the condition that defines a particle of this type is given by $\Psi=\Psi_{C}$
(i.e., $\Psi$ is invariant under charge conjugation). This is called
the Majorana condition. This condition imposed on the Dirac wave function
in the Majorana representation implies that this wave function must
be real. On the other hand, in the Majorana representation, the gamma
matrices must satisfy the relation $(\mathrm{i}\hat{\gamma}^{\mu})^{*}=\mathrm{i}\hat{\gamma}^{\mu}$,
which implies that the Dirac operator $\mathrm{i}\hat{\gamma}^{\mu}\partial_{\mu}-\frac{1}{\hbar c}(\mathcal{S}+\mathrm{m}c^{2})\hat{1}_{2}$
is real. Thus, in choosing the matrices in a Majorana representation,
the solutions of the Dirac equation can be chosen to be real but do
not need to be real. Thus, a Majorana particle cannot be described
only with the Dirac equation in the Majorana representation without
imposing the Majorana condition. 

Precisely, by considering four examples or distinct physical situations
(i.e., different borders and Lorentz scalar potentials), we constructed
real general solutions of the Dirac equation in (1+1) dimensions,
in the Majorana representation. The latter equation is a real system
of coupled equations that can describe a Majorana particle as long
as its solutions are real-valued. Clearly, the real general solutions
obtained here are a superposition of real solutions for a given label
or quantum number, but each of these real solutions is not a stationary
solution but a superposition of positive and negative energy complex
eigenstates (unless it is the real state corresponding to zero energy).
Likewise, these real general solutions can perfectly be square integrable. 

Because the equation describing the Majorana particle is the Dirac
equation with a Lorentz scalar potential -although with the restriction
imposed by the Majorana condition- the study of the supersymmetric
procedure in these circumstances turns out to be more important. In
particular, this task could give us new solutions that describe the
Majorana particle. A very good reference on this matter is Ref. \cite{RefV}. 

\begin{acknowledgments}
I thank Valedith Cusati, my wife, for all her support. Also,
I would like to express my gratitude to the reviewer for his/her interest,
comments and suggestions. 
\end{acknowledgments}

\end{document}